\documentstyle[aps,prl,epsfig]{revtex}
\begin{document}
\draft
\title{ Pattern Formation on Trees}
\author{M. G. Cosenza$^1$ and K. Tucci$^2$} 
\address{
$^1$Centro de Astrof\'{\i}sica Te\'orica;  $^2$SUMA-CESIMO \\Facultad de Ciencias,
Universidad de Los Andes,\\
 Apartado Postal 26 La Hechicera, M\'erida 5251, Venezuela}
\maketitle
\begin{abstract}
Networks having the geometry and the connectivity of trees are considered as the
spatial support of spatiotemporal dynamical processes. A tree is characterized by two
parameters: its ramification and its depth. The local dynamics at the nodes of a tree
is described by a nonlinear map, given rise to a coupled map lattice system. The
coupling is expressed by a matrix whose eigenvectors constitute a basis on which
spatial patterns on trees can be expressed by linear combination. The spectrum of
eigenvalues of the coupling matrix exhibit a nonuniform distribution which manifest
itself in the bifurcation structure of the spatially synchronized modes. These models
may describe reaction-diffusion processes and several other phenomena occurring on
heterogeneous media with hierarchical structure.
\end{abstract}
\vspace{1cm}
\pacs{ PACS Numbers: 05.45.+b, 02.50.-r}

\section{Introduction}

In general, media that support nonequilibrium pattern formation processes are
nonuniform on some length scales. Often the nonuniformity arises from the intrinsic
heterogeneous character of the medium, typical of pattern formation in many biological
systems, or from random discontinuities or from clustering in the medium. It is well
known that heterogeneities can have significant effects on the forms of spatial
patterns; for example they can produce reberberators in excitable media and defects
can serve as nucleation sites for domain growth processes. Recently, there has been
much interest in the study of dynamical processes on nonuniform networks, such as
fractal lattices \cite{CK1,CK3}, small world networks \cite{Watts}, hierarchically
interacting systems \cite{Gade1,Sinha}, random systems \cite{Gade2}, etc. An
especially interesting class of nonuniform geometries are trees whose hierarchical
structure and lack of translation symmetry can give rise to a number of distinctive
features in their dynamical and spatial properties. Examples of phenomena where
hierarchical networks appear include DLA clusters, capillarity, chemical reactions in
porous media \cite{Kopel}, turbulence \cite{Sree}, ecological systems \cite{Hogg},
interstellar cloud complexes \cite{Scalo}, etc. Hierarchical structures have also been
studied in neural nets, because of their exponentially large storage capacity
\cite{Admit}. Although many hierarchical structures found in nature have random
ramifications, here we study the case of simple, deterministic tree-like lattices.
This allows to focus on the changes induced in spatiotemporal patterns as a result of
the hierarchical structure of the interactions in the system.

In this article we consider discrete reaction-diffusion processes occurring on general
trees. The spatiotemporal dynamics corresponds to a coupled map lattice system defined
on the geometrical support of a tree. In Sec. II, the coupled map lattice models for
the study of general trees are introduced. A general notation for the treatment of
deterministic trees is also defined. The diffusion coupling among neighboring sites of
the lattice is described by a matrix, which exhibits an ordered structure. In Sec.
III, the spectrum of eigenvalues and eigenvectors of the coupling matrix. The
eigenvectors form a complete basis on which all spatial patterns can be expressed as a
linear combination. Explicit calculations of the eigenvalues and eigenvectors are
provided. Sec. IV presents a study of the bifurcation structure and stability of
spatially homogeneous, periodic patterns on trees for a local dynamics described by
the logistic map. Specific features emerge as a consequence of the ramified
character of the lattice. Finally, a discussion of the results is given in Sec. V.

\section{Coupled map lattice model}

Trees constitute a class of hierarchical networks which can be generated by a process
of successive branching starting from an initial element. In this paper, it is assumed
that the branching rule, or ramification number $R$, is the same through the network.
At the initial level of the tree, which we call level $0$, there is one element or
cell which splits into $R$ branches connecting $R$ daughters cells, which comprise the
level of construction $1$. Each cell then splits into $R$ daughters, producing $R^2$
sites at level $2$. This construction continues until some level $L$, which we call
the depth of the tree. There are $R^l$ cells at the level of construction, or layer,
$l$, where $l = 0,1,2,\dots,L$. Thus, each cell in the lattice has one parent and $R$
daughters, except for the level $0$ cell, which has no parent, and for boundary cells
at the level $l=L$, which have no daughters. The number of cells lying on the boundary
is $R^L$. The total number of cells on the tree, or the system size, is $N
=(R^{L+1}-1)/(R-1)$.

A cell belonging to a level $l$ is connected only to neighbor cells
lying in adjacent levels on the tree, i.e., to its parent and daughters.
We do not consider direct interactions among cells belonging to the
same level of construction. With these prescriptions, a tree is completely
characterized by two parameters: the ramification $R$ and the depth $L$.
We shall use the notation $(R,L)$ to indicate a tree possessing those
parameters.

Each cell in the tree can be specified by a sequence of symbols $(\alpha_1 \alpha_2
\dots \alpha_l)$, where $l$ is the level to which the cell belongs. A symbol
$\alpha_k$ can be take any value in a collection of $R$ different digits forming an
enumeration system in base $R$, which we denote by $\{\epsilon_1, \epsilon_2, \dots,
\epsilon_R\}$. The level $0$ cell can be assigned a different symbol, say $(\alpha_0)
= (\epsilon_0)$.

A cell identified with the sequence $(\alpha_1 \alpha_2 \dots \alpha_l)$, with $l>0$,
is always connected to its parent which has the label $(\alpha_1 \alpha_2 \dots
\alpha_{l-1})$, where the sequence of the $l-1$ symbols is the same as in the first
$(l-1)$ symbols of the daughter cell $(\alpha_1 \alpha_2 \dots \alpha_l)$. If $l<L$,
the cell $(\alpha_1 \alpha_2 \dots \alpha_l)$ is also connected to its $R$ daughters
which are labeled by $(\alpha_1 \alpha_2 \dots \alpha_l \epsilon_1), (\alpha_1
\alpha_2 \dots \alpha_l \epsilon_2), \dots, (\alpha_1 \alpha_2 \dots \alpha_l
\epsilon_R)$; where the sequence of the first $l$ symbols is the same as in the mother
cell $(\alpha_1 \alpha_2 \dots \alpha_l)$. Figure~1 shows a tree with ramification
number $R=3$ and depth $L=3$, indicating the labels on the cells.

A tree might be considered as the spatial support of spatiotemporal dynamical
processes with either discrete or continuous time  We focus on reaction-diffusion and
pattern formation phenomena on trees. The dynamical systems considered here are
defined by associating a nonlinear function with each cell of a given tree and
coupling these functions through nearest-neighbor diffusion interaction. In this way,
a coupled map lattice describing a reaction-diffusion dynamics on a tree with
ramification $R$ and depth $L$ can be expressed as

\begin{enumerate}
\item Level $0$ cell:
\begin{equation}
\label{eq1}
 x_{t+1}(\epsilon_0) = f(x_t(\epsilon_0)) + \gamma\left[\sum_{i=1}^{R}
x_t(\epsilon_i) - R x_t(\epsilon_0)\right];
\end{equation}

\item Level $0<l<L$ cells:
\begin{equation}
\label{eq2}
\begin{array}{l}
x_{t+1}(\alpha_1 \dots \alpha_l) = f(x_t(\alpha_1 \dots \alpha_l)) + \cr + \gamma
\left[ \sum\limits_{i=1}^{R} x_t(\alpha_1 \dots \alpha_l \epsilon_i) + x_t(\alpha_1
\dots \alpha_{l-1}) - (R+1) x_t(\alpha_1 \dots \alpha_l)\right];
\end{array}
\end{equation}

\item Level $l=L$ cells:
\begin{equation}
\label{eq3}
 x_{t+1}(\alpha_1 \dots \alpha_L) = f(x_t(\alpha_1 \dots \alpha_L)) +
\gamma \left[x_t(\alpha_1 \dots \alpha_{L-1}) - x_t(\alpha_1 \dots \alpha_L) \right];
\end{equation}
\end{enumerate}
where $x_t(\alpha_1 \dots \alpha_l)$ gives the state of the cell labeled by
$(\alpha_1, \dots, \alpha_l)$ on the tree at discrete time $t$; $f(x_t)$ is a
nonlinear function specifying the local dynamics; and $\gamma$ is a parameter that
measures the strength of the coupling among neighboring cells and plays the role of a
homogeneous diffusion constant. This type of coupling is usually called backward
diffusive coupling and corresponds to a discrete version of the Laplacian in
reaction-diffusion equations.

The above coupled map lattice equations can be generalized to include other coupling
schemes, non-uniform coupling, varying ramifications within the network, or
continuous-time local dynamics. Different spatiotemporal phenomena can be investigated
on tree-like structures by providing appropriate local dynamics and couplings.

Equations (\ref{eq1})-(\ref{eq3}) can be written in vector form as

\begin{equation}
\label{vector}
{\bf x}_{t+1} = {\bf f}({\bf x}_t) + \gamma{\bf M}{\bf x}_t.
\end{equation}

The state vector ${\bf x}_t$ possesses $N$ components $x_t(j), j=0,\dots,N-1$,
corresponding to the states $x_t(\alpha_1 \dots \alpha_l)$ of the cells on a tree
$(R,L)$ labeled with the sequence of symbols $(\alpha_1 \dots \alpha_l)$. The matrix
${\bf M}$ expresses the coupling among the components $\{x_t(j)\}$.

The components of ${\bf x}_t$ may be ordered as follows. The level $0$ cell is
assigned the index $j=0$. All other cells labeled by $(\alpha_1 \dots \alpha_l)$ can
be associated to a unique integer index $ j=1,\dots,N-1$, by the rule
\begin{equation}
(\alpha_1 \dots \alpha_l) \longleftrightarrow j=\frac{R^{l} - 1}{R-1} + \sum_{k=1}^{l}
\alpha_k R^{l-k}.
 \label{eq:rule}
\end{equation}
As an example, consider the cell labeled by the sequence $(21)$ on the tree of Fig.
(1). In this case $R=3$ and the cell belongs to the level $l=2$. According to the rule
of Eq.(\ref{eq:rule}), its index is $j=11$. Its parent is labeled by $(2)$, and has
the index $j=3$; while its daughters, labeled by $(210)$, $(211)$ and $(212)$, are
assigned indexes $j=34, 35$ and $36$, respectively.

Given this notation, the $j-$component of the vector-valued function ${\bf f}({\bf
x}_t)$ is $[{\bf f}({\bf x}_t)]_j = f(x_t(j))$. For a tree characterized by $(R,L)$,
the elements of its corresponding coupling matrix ${\bf M}$, denoted by $M(i,j)$,
$(i,j = 0,1,\dots, N-1)$, are

\begin{equation}
M(i,j) = M(j,i) = \cases{ -R &, if $i=0$ and $j=0$ \cr
    -(R+1)&, if $i=j$      and $0<i<\frac{R^L-1}{R-1}$ \cr
    -1     &, if $i=j$      and $i\geq\frac{R^L-1}{R-1}$\cr
    1     &, if $i \neq j$ and $\left( j=\mbox{int}(\frac{i-1}{R}) \; \mbox{or} \;
                           i=\mbox{int}(\frac{j-1}{R}) \right) $\cr
    0     &, elsewhere,
}
\end{equation}
where $\mbox{int}(q)$ means the integral part of $q$. The matrix ${\bf M}$ is a $N
\times N$ real and symmetric matrix. It should be emphasized that ${\bf M}$ plays the
role of a spatially discrete diffusion operator on treelike networks, analogous to the
Laplacian in a spatially continuous reaction-diffusion equation.

\section{Spectrum of the coupling matrix}

Similarly to reaction-diffusion processes on fractal lattices \cite{CK1}, the spatial
patterns that can arise on trees are determined by the eigenmodes of the coupling
matrix ${\bf M}$. Additionally, the stability of the synchronized states is related to
the set of eigenvalues of ${\bf M}$.

In order to analyze the eigenvector problem, consider a tree with ramification $R$ and
depth $L$ on which a spatiotemporal dynamics has been defined in the vector form of
Eq.(\ref{vector}). The complete set of orthonormal eigenvectors of the corresponding
matrix ${\bf M}$ can be described as the superposition of two distinct subsets of
eigenmodes. One subset, which will be denoted by $\{{\bf u}_n\}$, comprises those
eigenvectors associated to non-degenerate eigenvalues; and the other subset contains
the eigenvectors corresponding to degenerate eigenvalues of ${\bf M}$, and will be
represented by $\{{\bf v}_{m s}^g\}$ (the indices refer to the degeneracy, as
explained bellow). Thus, the complete set of eigenvectors of ${\bf M}$ is $\{{\bf
u}_n\} \cup \{{\bf v}_{m s}^g\}$. Each eigenvector describes a basic spatial pattern
that may take place on a tree characterized by $(R,L)$.

\subsection*{Non-degenerate eigenmodes}

The eigenvectors of ${\bf M}$ belonging to the non-degenerate subset $\{{\bf u}_n\}$
satisfy

\begin{equation}
\label{ev1}
 {\bf M} {\bf u}_n = b_n {\bf u}_n ,  \quad\quad n=1,2,\dots,\nu;
\end{equation}
where $b_n$ is the eigenvalue associated to the eigenvector ${\bf u}_n$. There are
$\nu$ distinct eigenvectors with their corresponding eigenvalues in this subset. The
j-component of a vector ${\bf u}_n$ is $[{\bf u}_n]_j =
u_n(\alpha_1,\alpha_2,\dots,\alpha_l)$, according to the associating index rule,
Eq.(\ref{eq:rule}). Any eigenvector ${\bf u}_n$ in this subset is characterized by the
following property: all its components corresponding to cells of the tree lying on the
same level or layer $l$ are identical, i.e.,
\begin{equation}
u_n(\alpha_1\alpha_,\dots\alpha_l) = u_n(\beta_1\beta_2\dots\beta_l),
\label{eq:identical layer}
\end{equation}
where $(\alpha_,\alpha_2\dots\alpha_l)$ and $(\beta_1\beta_2\dots\beta_l)$ label any
two cells belonging to the level $l$. A non-degenerate eigenvector of ${\bf M}$ thus
possesses homogeneous layers. Because of property (\ref{eq:identical layer}), we also
refer to the elements of $\{{\bf u}_n\}$ as {\it layered} eigenvectors.

An eigenvector ${\bf u}_n$ representing a tree of depth $L$ has $(L+1)$ levels,
including the level $0$ cell. Because of the homogeneous layer property, there can be
$\nu=L+1$ distinct eigenvectors in the subset $\{{\bf u}_n\}$ satisfying this
condition; one eigenvector for each homogeneous layer that can be formed. Note that
the number of different layered eigenvectors depends only on the depth $L$ of the tree
and not on its ramification $R$. In particular, the spatially homogeneous eigenmode of
${\bf M}$, which we denote by ${\bf u}_1$, belongs to the subset $\{{\bf u}_n\}$ and
its $N$ components are
\begin{equation}
u_1(\alpha_1\alpha_2\dots\alpha_l) = N^{-1/2}; \qquad \forall l,\quad \forall \alpha_k
\, \label{homoeigenvectorcomp}
\end{equation}
and
\begin{equation}
{\bf u}_1= \frac{1}{\sqrt{N}} \, \mbox{col}\,(1,1,\ldots,1).
\end{equation}
Since the eigenvectors of ${\bf M}$ are mutually orthogonal, all other eigenmodes, in
either subset $\{{\bf u}_n\}$ or $\{{\bf v}_{m s}^g\}$, must satisfy
\begin{equation}
\begin{array}{l}
    \sum\limits_{\alpha_1,\alpha_2,\dots,\alpha_l}
    u_n(\alpha_1\dots\alpha_l) = 0; \qquad
    n \neq 1,\quad \forall l,\quad \forall \alpha_k;
\cr
    \sum\limits_{\alpha_1,\dots,\alpha_l}
    v_{m s}^g(\alpha_1\dots\alpha_l) = 0; \qquad
    \forall l,\quad \forall \alpha_k;
\end{array}
\label{eq:orthogolity}
\end{equation}
that is, the sum of the components of any other eigenvector of
${\bf M}$ different that ${\bf u}_1$ is zero.

Figure 2(a) shows the three layered eigenvectors and their associated eigenvalues of the
coupling matrix corresponding to a tree with $R=3$ and $L=2$.

The set of eigenvalues arising from Eq.~(\ref{ev1}) may be ordered by decreasing value
as $\{b_1,b_2,\ldots,b_L,b_{L+1}\}$. By Gershgorin's theorem \cite{Book}, the
homogeneous eigenvector possesses the largest eigenvalue of ${\bf M}$, which is
$b_1=0$. On the other hand, for large $L$ the smallest eigenvalue is found to be
\begin{equation}
\lim_{L\rightarrow\infty} b_{L+1} = \frac{-(2+3R) - \sqrt{R(4+R)}}{2},
\end{equation}
and similarly, we find
\begin{equation}
\lim_{L\rightarrow\infty} b_{2} = \frac{-(R+2) + \sqrt{R(4+R)}}{2} \, .
\end{equation}

The eigenvalues $\{b_1,b_2,\ldots,b_L,b_{L+1}\}$ appear in pairs $b_n$ and $b_{n'}$,
as $b_{L+1}$ and $b_2$ above, according to the sign of the square root term. These
pairs are related by
\begin{equation}
\label{bn}
 b_n + b_{n'} = -2(R+1),
\end{equation}
where
\[
n + n' = \cases{\frac{L+3}{2} & , if $L$ is odd \cr
        L+3           & , if $L$ is even.}
\]

The eigenvalue $b_{(L+3)/2}$ arises whenever $L$ is odd, and it is not associated with
another $b_n$. Its value is
\begin{equation}
b_{\frac{L+3}{2}} = -(R+1). \label{eq:b_L+3/2}
\end{equation}

Thus, because of (\ref{bn}) and (\ref{eq:b_L+3/2}) the eigenvalues associated to the
non-degenerate eigenvectors satisfy
\begin{equation}
\sum_{n=1}^\nu b_n =
                -L(R+1).
\end{equation}

Figure 3 shows the spectrum of eigenvalues $\{ b_n\}$, indicated by black dots, for a
tree with ramification $R=3$ at successive depths $L$. Eigenvalues associated to
degenerate eigenvectors of ${\bf M}$, to be discussed next, are also shown in Fig. 3.

\subsection*{Degenerate eigenmodes}

The subset of degenerate eigenvectors $\{{\bf v}_{m s}^g\}$ of the coupling matrix
${\bf M}$ corresponding to a tree characterized by $(R,L)$ satisfy
\begin{equation}
{\bf M}{\bf v}_{m s}^g = a_{m s} {\bf v}_{m s}^g
\label{eq:16}
\end{equation}
where $a_{m s}$ is the eigenvalue associated to a group of $\delta$ degenerate
eigenvectors $\{{\bf v}_{m s}^1, {\bf v}_{m s}^2, \dots, {\bf v}_{m s}^\delta\}$
belonging to $\{{\bf v}_{m s}^g\}$. The index $g$ goes from $1$ to $\delta$ and counts
the different eigenvectors associated to the degenerate eigenvalue $a_{m s}$. The
integer indices $m$ and $s$ label different eigenvalues $a_{m s}$.

The $j-$component of a vector $\{{\bf v}_{m s}^g\}$ corresponds to a cell of the tree
labeled by the rule (\ref{eq:rule}), i.e., $[{\bf v}_{m s}^g]_j = {\bf v}_{m
s}^g(\alpha_1, \alpha_2, \dots, \alpha_l)$. The eigenmodes $\{{\bf v}_{m s}^g\}$ are
characterized by the following two properties,
\begin{equation}
{\bf v}_{m s}^g (\alpha_1 \alpha_2 \dots \alpha_l) = 0;\qquad l=0,1,2,\dots,L-m;
\label{eq:17}
\end{equation}
that is, all the components of ${\bf v}_{m s}^g$ lying in successive levels vanish up
to the level $l=L-M$; and
\begin{equation}
\sum_{\alpha_1 \alpha_2 \dots \alpha_l} {\bf v}_{m s}^g (\alpha_1 \alpha_2 \dots
\alpha_l) = 0; \label{eq:sum_property}
\end{equation}
i.e., the sum of the components of an eigenvector ${\bf v}_{m s}^g$, lying on the same
level of the tree spatially described by ${\bf v}_{m s}^g$, is zero.

An eigenvector ${\bf v}_{m s}^g$ is spatially uniform in part, having all its
components, or equivalent cells, equal to zero up to level $L-m$. The index $m$ counts
the number of remaining non-vanishing layers in the eigenvector ${\bf v}_{m s}^g$, and
its possible values are $m = 1, 2, \dots, L$. Each of the $m$ non-vanishing layers may
be homogeneous, but different among each other. The index $s$ counts the number of
possible different eigenvectors with $m$ different homogeneous, non-vanishing last
layers. Thus, $s$ may take the values $s = 1, 2, \dots, m$.

The index $g$ lifts the degeneracy of vectors with the same indices $m$ and $s$. The
remaining, non-vanishing last layers may in fact be non-homogeneous, and may consists
of subtrees with homogeneous sub-levels, which would reproduce the structure of the
layered non-degenerate eigenvectors ${\bf u}_n$. The level $l=L-m$ is the last
vanishing layer in an eigenvector having $m$ non-vanishing layers. On this layer,
there are $R^{L-m}$ components or cells, and each of these cells gives origin to $R$
layered subtrees, i.e., subtrees with homogeneous layers. These $R$ subtrees
themselves are related by the sum property Eq.~(\ref{eq:sum_property}), which results
in $(R-1)$ linearly independent subtrees. Therefore, the number of linearly
independent eigenvectors ${\bf v}_{m s}^g$ with the same values $m$ and $s$ is
$\delta=(R-1)R^{L-m}$. The index $g$ expresses the degeneracy of the eigenvectors
associated to the eigenvalue $a_{m s}$, and it may take the values $g = 1, 2, \dots,
(R-1)R^{L-m}$. In this way, the set of degenerate eigenvectors $\{{\bf v}_{m s}^g\}$
of the matrix ${\bf M}$ corresponding to a tree $(R,L)$ is fully described.

Figure 2(b) shows the subset of degenerate eigenvectors $\{{\bf v}_{m s}^g\}$ and the
eigenvalues corresponding to a tree characterized by $(R,L)=(3,2)$.

The index $m$ also expresses the form in which an eigenvalue $a_{m s}$ arises in the
spectrum of eigenvalues of ${\bf M}$. An eigenvalue $a_{m s}$ appears for the first
time at a level $l=m>1$ and stays in the spectrum of eigenvalues of ${\bf M}$ at
subsequent levels up to $l=L$. Thus the index $m$ may take the values $m=1, 2, \dots,
L$. On the other hand, $m$ different eigenvalues arise at the level of construction
$l=m$ of the tree, which are counted by the index $s = 1, 2, \dots, m$.

The total number of distinct eigenvalues of type $a_{m s}$ belonging to the spectrum
of a matrix ${\bf M}$ associated to a tree $(R,L)$ is
\begin{equation}
\sum_{m=1}^L m = \frac{L(L+1)}{2}.
 \label{ams}
\end{equation}
Additionally, the $m$ eigenvalues $a_{m s}$ that appear at a level $l=m$ satisfy the
following property
\begin{equation}
\sum_{s=1}^m a_{m s} = -(m-1)(R+1) - 1,
\end{equation}
and therefore the total sum of eigenvalues $a_{m s}$ for a matrix ${\bf M}$ associated
to a tree $(R,L)$ will be,
\begin{displaymath}
\sum_{m=1}^L \sum_{s=1}^m a_{m s} = -\sum_{m=1}^L (m-1)(R+1) - L
\end{displaymath}
\begin{equation}
= -\frac{(R+1)(L-1)L}{2} - L .
\end{equation}

Figure 3 shows the spectrum of eigenvalues $\{a_{m s}\}$ for a tree with ramification
$R=3$ at successive depths $L$. Eigenvalues $\{ b_n\}$ corresponding to non-degenerate
eigenvectors are also shown there. Thus the distribution of the full spectrum of
eigenvalues of the coupling matrix ${\bf M}$ can be seen as a function of the depth of
the tree in Fig. 3. Note that the full spectrum $\{b_n\} \cup \{a_{m s}\}$ is always
contained between the eigenvalues $b_1=0$ and $b_{L+1}$.

The total number of distinct eigenvalues of ${\bf M}$, including both types $a_{ms}$
and $b_n$, and denoted by $\Omega$, is
\begin{equation}
 \label{numeigenvalues}
\Omega=(L+1) + \frac{L(L+1)}{2} = \frac{(L+1)(L+2)}{2}.
\end{equation}

Note that the total number of eigenvalues in the spectrum of ${\bf M}$ is determined
only by the depth of the tree and is independent of its ramification, although the
specific values of the eigenvalues do depend on both $R$ and $L$. Since there appear
$m$ eigenvalues of type $a_{m s}$ at each level $l=m$ and there are $(R-1)R^{L-m}$
degenerate eigenvectors ${\bf v}_{m s}^g$ associated to each eigenvalue $a_{m s}$, the
total number of independent eigenmodes in the subset $\{{\bf v}_{m s}^g\}$ is
$\sum_{m=1}^L m(R-1)R^{L-m}$. In the non-degenerated subset $\{{\bf u}_n\}$ there are
$(L+1)$ independent eigenvectors, as we saw before. Therefore, the total number of
independent eigenmodes of ${\bf M}$ is
\begin{displaymath}
(L+1) + \sum_{m=1}^L m(R-1)R^{L-m} =
\end{displaymath}
\begin{displaymath}
=(L+1) + (R-1)R^L \; \frac{R^{L+1} - R(L+1) + L}{R^L(R-1)^2}=
\end{displaymath}
\begin{equation}
 =\frac{R^{L+1} - 1}{R - 1}=N,
\end{equation}
as expected.

Figure 4(a) shows the complete spectrum of eigenvalues of ${\bf M}$ and the degeneracy
fraction of each eigenvalue, for a tree characterized by parameters $(R,L)=(2,11)$.
The $L+1=12$ eigenvalues $b_n$ are indicated by dots and they are non-degenerate,
while the degeneracy $\delta=2^{11-m}$ of each of the $L(L+1)/2=66$ eigenvalues
$a_{ms}$ is plotted as a vertical bar. It is evident that both the distribution of
eigenvalues and their degeneracies are nonuniform. Another convenient representation
of the scaling properties of the spectrum of eigenvalues of the coupling  matrix can
be obtained by plotting the accumulated sum of the degeneracies of all eigenvalues,
that is the measure of the spectrum of ${\bf M}$ (denoted by $\rho$), on the
eigenvalue axis for large $L$, as in Fig. 4(b). The resulting graph presents the
features of a devil´s staircase, a fractal curve arising in a variety of nonlinear
phenomena.

The eigenmodes of the coupling matrix reflect the topology of the tree and they are
analogous to the Fourier eigenmodes appearing in regular Euclidean lattices. In this
sense, conditions Eq. (\ref{eq:identical layer}) and
Eqs.~(\ref{eq:17})-(\ref{eq:sum_property}) represent different wavelengths on a tree
characterized by parameters $(R,L)$.

\subsection*{An example}

As an example of calculation of eigenvectors, consider any tree with ramification $R$
and depth $L=1$. The number of cells of the tree is $N=(R^2 - 1)/(R - 1) = R+1$. Thus,
the tree consists of a mother cell at level $l=0$ connected to its $R$ daughters at
level $l=1$. The corresponding $(R+1)\times(R+1)$ coupling matrix has the form
\begin{equation}
{\bf M} = \left(\begin{array}{c c c c c c}
-R & 1 & 1 & 1 & \dots & 1 \cr
1 & -1 & 0 & 0 & \dots & 0 \cr
1 & 0 & -1 & 0 & \dots & 0 \cr
1 & 0 & 0 & -1 & \dots & 0 \cr
\vdots & \vdots & \vdots & \vdots & \ddots & \vdots \cr
1 & 0 & 0 & 0 & \dots & -1
\end{array}\right) \, ,
\label{}
\end{equation}
and the associated eigenvectors of ${\bf M}$ have $(R+1)$ components. There exist two
non-degenerate eigenvectors, which are the homogeneous ${\bf u}_1 =
\frac{1}{\sqrt{R+1}}\, \mbox{col}\,(1,1,\dots,1)$, and ${\bf u}_2$, associated to the
eigenvalues $b_1=0$ and $b_2$, respectively. There is only the eigenvalue $a_{1 1}$
associated to $\delta=(R-1)$ degenerate eigenvectors $\{{\bf v}_{1 1}^1, {\bf v}_{1
1}^2, \dots, {\bf v}_{1 1}^{R-1}\}$. The total number of independent eigenmodes of
${\bf M}$ is $(R+1)$, and the total number of distinct eigenvalues is $\Omega=3$, in
agreement with Eq.(\ref{numeigenvalues}).

All the eigenvectors of ${\bf M}$ have the level $l=0$, with one component or cell;
and the level $l=1$, for which there are $R$ components. The layered eigenvector ${\bf
u}_2$ will have the form ${\bf u}_2 = \mbox{col}(x, y, y, \dots, y)$. Its eigenvalue
equation ${\bf M}{\bf u}_2 = b_2{\bf u}_2$, plus the normalization condition $|{\bf
u}_2| = 1$, yield the relations
\begin{equation}
\begin{array}{r c l}
-Rx + Ry   & = & b_2x, \cr
x - y      & = & b_2y, \cr
x^2 + Ry^2 & = & 1,
\end{array}
\end{equation}
whose solutions are $b_2 = -(R+1)$,  $x=\frac{-R}{\sqrt{R(R+1)}}$,
$y=\frac{1}{\sqrt{R(R+1)}}$. Thus,
\begin{equation}
{\bf u}_2= \frac{1}{\sqrt{R(R+1)}} \, \mbox{col} \,(-R, 1, 1, \ldots, 1).
\end{equation}

On the other hand, the degenerate eigenmode ${\bf v}_{1 1}^1$ has the form ${\bf v}_{1
1}^1 = \mbox{col}\,(0, x_1, x_2, \dots, x_R)$, satisfying properties (\ref{eq:17}) and
(\ref{eq:sum_property}), as well as the eigenvector equation ${\bf M}{\bf v}_{1 1}^g =
a_{1 1}{\bf v}_{1 1}^g$ and the normalization condition $|{\bf v}_{1 1}^g| = 1$. These
relations lead to
\begin{equation}
\begin{array}{r c l}
\sum_{i=1}^R x_i   & = & 0, \cr
-x_i               & = & a_{1 1}x_i, \cr
\sum_{i=1}^R x_i^2 & = & 1,
\end{array}
\end{equation}
which imply that $a_{1 1} = -1$. Making $x_1 = x$ and $x_2=x_3=\dots=x_R=y$; we get
\begin{equation}
\begin{array}{r c l}
x+(R-1)y     & = & 0, \cr
x^2+(R-1)y^2 & = & 1,
\end{array}
\end{equation}
with solutions, $x=\frac{-(R-1)}{\sqrt{R(R-1)}}$, $y=\frac{1}{\sqrt{R(R-1)}}$. Thus
\begin{equation}
{\bf v}_{1 1}^1 = \frac{1}{\sqrt{R(R-1)}} \,\mbox{col} \left( 0, -(R-1), \underbrace{
  1,\dots,1}_{(R-1) \, \mbox{times}} \right) \, .
\end{equation}

For ${\bf v}_{1 1}^2 = \mbox{col}\, (0, 0, x_2, \dots, x_R)$ the procedure can be
repeated by making $x_2=z$, $x_3=x_4=\dots=x_R=w$, obtaining
\begin{equation}
{\bf v}_{1 1}^2 = \,\mbox{col} \,\left(0,0,z, \underbrace{w,\dots,w}_{(R-2) \,
\mbox{times} } \right),
\end{equation}
\begin{displaymath}
\Longrightarrow \begin{array}{r c l}
    z+(R-2)w       & = & 0,\cr
    z^2 + (R-2)w^2 & = & 1,
\end{array}
\end{displaymath}
\begin{displaymath}
z=\frac{-(R-2)}{\sqrt{(R-1)(R-2)}}, \quad w=\frac{1}{\sqrt{(R-1)(R-2)}} \, .
\end{displaymath}

In general, we get
\begin{equation}
{\bf v}_{1 1}^g = \mbox{col} \left( \underbrace{0,\dots,0}_{g \, \mbox{times}},z,
\underbrace{w,\dots,w}_{(R-g) \, \mbox{times}} \right),
\end{equation}
\begin{displaymath}
\Longrightarrow \begin{array}{r c l}
    z+(R-g)w       & = & 0,\cr
    z^2 + (R-g)w^2 & = & 1,
\end{array}
\end{displaymath}
\begin{displaymath}
z=\frac{-(R-g)}{\sqrt{(R-g+1)(R-g)}}, \quad w=\frac{1}{\sqrt{(R-g+1)(R-g)}}, \quad
g=1,2,\dots,(R-1);
\end{displaymath}
giving $(R-1)$ eigenvectors in the degenerate subset $\{{\bf v}_{m s}^g\}$ for this
example. With the addition of the two non-degenerate eigenvectors ${\bf u}_1$ and
${\bf u}_2$, there are $N=R+1$ independent eigenvectors. Therefore, all the
eigenvectors and eigenvalues associated to a matrix ${\bf M}$ corresponding to a tree
characterized by parameters $(R,1)$ are accounted for. Note that since the eigenvalue
$a_{11} = -1$ appears at level $l=m=1$, it will stay in the spectrum of eigenvalues of
${\bf M}$ for all subsequent levels of construction, i.e., $a_{11}=-1$ arises for
trees of any depth. Similarly, the eigenvalue $b_1=0$ and its associated homogeneous
eigenvector ${\bf u}_1$ always appear in a tree.

\section{Bifurcation structure and stability of spatially synchronized states}

Spatially synchronized states in extended systems are relevant since we are often
interested in the mechanism by which a uniform system breaks its symmetry to form a
spatial pattern as a parameter is changed. Consider spatially synchronized, period $K$
states such as $x_t(\alpha_1, \ldots,\alpha_l)= \bar{x}_k$, $\forall (\alpha_1,
\ldots,\alpha_l)$; where $\bar{x}_k$ ,$(k=1,2,\ldots,K)$ is a period $K$ orbit of the
the local map, $f^{(K)}(\bar{x}_k)=\bar{x}_k$. The linear stability analysis of
periodic, synchronized states in coupled map lattices is carried out by the
diagonalization of  ${\bf M}$ in Eq. (\ref{vector}), and it leads to the conditions
\cite{Waller}
\begin{equation}
\label{stability}
\prod_{k=1}^K \left[ f'(\bar{x}_k) + \gamma \mu \right]= \pm 1,
\end{equation}
where $\mu$ is an eigenvalue, in either set $\{ b_n\}$ or  $\{a_{m s}\}$, of the
coupling matrix ${\bf M}$ describing a tree $(R,L)$. There are $\Omega=(L+1)(L+2)/2$
different values of $\mu$ (Eq. (\ref{numeigenvalues})) to be used in Eqs.
(\ref{stability}).

The nonuniform distribution of the eigenvalue spectrum is manifested in the stability
of the synchronized states through this relation and give rise to important
differences when compared, for instance, with the bifurcation structure on regular
lattices. As an application, consider a local dynamics described by the logistic map,
$f(x)=\lambda x (1-x)$. In this case, the bifurcation conditions, Eq.
(\ref{stability}), for the period $K=2^p$, synchronized state on a tree characterized
by parameters $(R,L)$ can be expressed as the set of curves
\begin{equation}
\label{synlog}
 S_L^p(\mu) \equiv \prod_{k=1}^{2^p} \left[ \lambda(1-2\bar{x}_k) + \gamma \mu
\right]= \pm 1.
\end{equation}
For each sign, Eqs.~(\ref{synlog}) yield  $(L+1)(L+2)/2$ boundary curves in the plane
$(\gamma,\lambda)$ which determine the stability regions of the period $2^p$,
synchronized states on the tree.

The scaling structure for the period-$2^p$, synchronized states in trees is similar to
that of a any lattice described by a diffusive coupling matrix, since the form of Eq.
(\ref{synlog}) is the same in any case. As for any lattice  (for example regular
Euclidean lattices \cite{Waller} or fractal lattices \cite{CK1}), the stability
regions for the period-$2^p$, synchronized states in the $(\gamma, \lambda)$ plane
scale as $\lambda \sim \delta^{-p}$, and $\gamma \sim \alpha^{-p}$, where
$\delta=4.669\ldots$ and $\alpha=-2.502\ldots$ are Feigenbaum's scaling constants for
the period doubling transition to chaos. However, the specific structure of the
eigenvalue spectrum of the coupling matrix determines the shapes and gaps of the
regions of stability of synchronized, periodic states.

The boundary curves Eqs. (\ref{synlog}) for the synchronized, fixed point state
$(p=0)$ are given by the straight lines
\begin{equation}
\lambda=\mu \gamma +1, \quad\quad\quad \lambda=\mu \gamma +3;
\end{equation}
which are first crossed for the most negative eigenvalue, $\mu=b_{L+1}$. Fig.  5(a)
shows the boundary curves Eq. (\ref{synlog}) in the plane $(\gamma, \lambda)$ for the
period-two ($p=1$), synchronized state on a tree with $(R,L)=(3,3)$, which are given
by the two sets
\begin{equation}
\label{boun1} S_3^1(b_n)= -\lambda^2+2\lambda+4+\gamma b_n(\gamma b_n-2)= \, \pm 1 ;
\quad\quad
 n=1,2,3,4 ;
\end{equation}
and
\begin{eqnarray}
\label{boun2} S_3^1(a_{ms})=
 -\lambda^2+2\lambda+4+\gamma a_{ms}(\gamma a_{ms}-2)= &\pm
1 ;\\
 m=1,2,3;  \quad s=1,\ldots,m. &  \nonumber
 \end{eqnarray}
The boundary between the synchronized, fixed point state and the synchronized
period-two state is at $\lambda=3$. The upper boundaries (corresponding to $-1$ in the
r.h.s of Eqs. (\ref{boun1})-({\ref{boun2})) have minima $\lambda_{\min}=1+\sqrt{5}$ at
values $\gamma_{\min}=1/b_n$ and $\gamma_{\min}=1/a_{ms}$ (for any period $2^p$,
$\lambda_{\min}$ depends on $p$). Fig. 5(b) shows a magnification of Fig. 5(a) around
the minima of the upper boundaries. The distribution of the minima $\gamma_{\min}$ and
the presence of nonuniformly distributed gaps (niches) in the boundary curves reflect
the nonuniform structure of the eigenvalue spectrum. Since the nonuniformity in the
distribution of eigenvalues persists at any depth $L$ of a tree, this property allows
for regions of stability of the synchronized states (niches) characteristic of trees
and which are not present in other geometries, for example in regular lattices, where
the distribution of eigenvalues of the coupling matrix is uniform and continuous in
the limit of infinite size lattices.

The set of eigenvectors $\{{\bf u}_n\} \cup \{{\bf v}_{m s}^g\}$ of the coupling
matrix ${\bf M}$ constitute a complete basis (normal modes) on which a state ${\bf
x}_t$ of the system can be represented as a linear combination of these vectors. The
evolution of ${\bf x}_t$ then reflects the stabilities of the normal modes. Fig. 5(b)
shows how the synchronized state may become unstable through crossing of the upper
boundary; the first boundary segment crossed determines the character of the
instability. For example, consider an initial state consisting of a small perturbation
of the synchronized, period-$2$ state at parameter values just beyond the boundary
segment corresponding to $a_{2 2}$ indicated by a cross in Fig. 5(b), where this
initial state is unstable. The inhomogeneous period-$4$ final spatial pattern is
represented in Fig. 6; it corresponds to a linear combination of the six eigenmodes
$\{{\bf v}_{2 2}^g; \, g=1,\ldots,6 \}$ and associated to the degenerate eigenvalue
$a_{2 2}$ of the matrix ${\bf M}$ corresponding to the tree $(R,L)=(3,3)$. All other
modes are unstable in this region of parameter space. For any depth $L$ of the tree,
and any period $2^p$, the boundary curve $S_3^1(a_{2 2})=-1$ separates a niche of the
synchronized state from the stable region for the eigenmodes ${\bf v}_{2 2}^g$
corresponding to $a_{2 2}$. Thus, a transition between these two spatial patterns can
always be observed in the appropriate regions of the $(\gamma, \lambda)$ plane.

\section{Conclusions}
In a system of interacting agents, such as the models presented in this article, the
coupling matrix contains the connectivity of the network and it determines the spatial
patterns that can arise in the system. The underlying inhomogeneous structure of trees
has pronounced effects on the spatial patterns that can be formed by
reaction-diffusion processes on these lattices. The spatial patterns are determined by
the eigenvectors of the coupling matrix ${\bf M}$; and the stability of the
synchronized states is determined by the corresponding eigenvalues. The set of normal
modes of the coupling matrix reflect the connectivity of the tree. These modes have
complex spatial forms but they are analogous to the Fourier eigenmodes arising in
regular Euclidean lattices. On the other hand, the distribution of eigenvalues of
${\bf M}$ and their degeneracies are nonuniform. These features affect the bifurcation
properties of dynamical systems such as coupled map defined on trees. The scaling
structure of the synchronized, period-doubled states is similar for both uniform and
hierarchical lattices, but the nature of the bifurcation boundaries is different. For
trees, the boundary curves are determined by the spectrum of eigenvalues of the
coupling matrix, which has a nonuniform density. The nonuniform distribution of
eigenvalues leads to gaps or niches in the boundary curves that are not present for
coupled maps on uniform lattices, where the spectrum of eigenvalues is continuous.

We have examined only the simplest spatiotemporal patterns that can be formed on
treelike geometries; however, the formalism presented in Sec.~II can be applied to
many other processes, such as nontrivial collective behavior, excitation waves, phase
transitions, domain segregation and growth on trees. The formalism is also useful for
continuous-time local dynamics. Similarly, extensions of this work are possible in
order to include networks with variable ramification and/or depths.

The study of dynamical systems defined on trees and other nonuniform substrates should
allow us to gain insight into previously unexplored spatiotemporal phenomena on
inhomogeneous systems and to understand better the relationship between topology and
collective properties of networks.

\section*{Acknowledgment}
This work was supported by the Consejo de Desarrollo
Cient\'{\i}fico, Human\'{\i}stico y Tecnol\'ogico of
the Universidad de
Los Andes, M\'erida, Venezuela.

\begin{center}
\begin{figure}
\epsfig{file=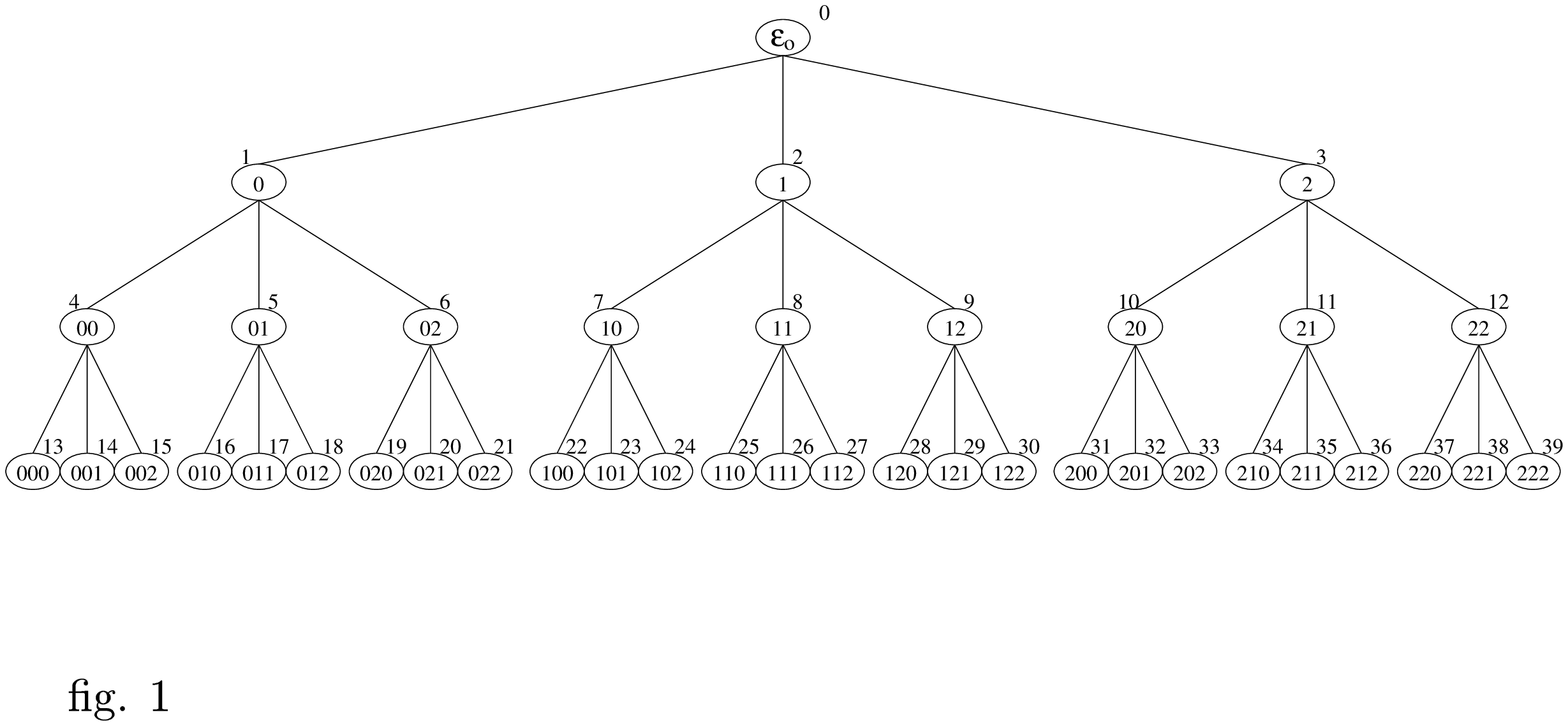, width=1.0\textwidth ,angle=90, clip=}
\caption{Tree with ramification $R=3$ and depth $L=3$, showing the labels on the cells.
The corresponding vector-component index $j$ is indicated besides each cell.}
\end{figure}
\begin{figure}
\epsfig{file=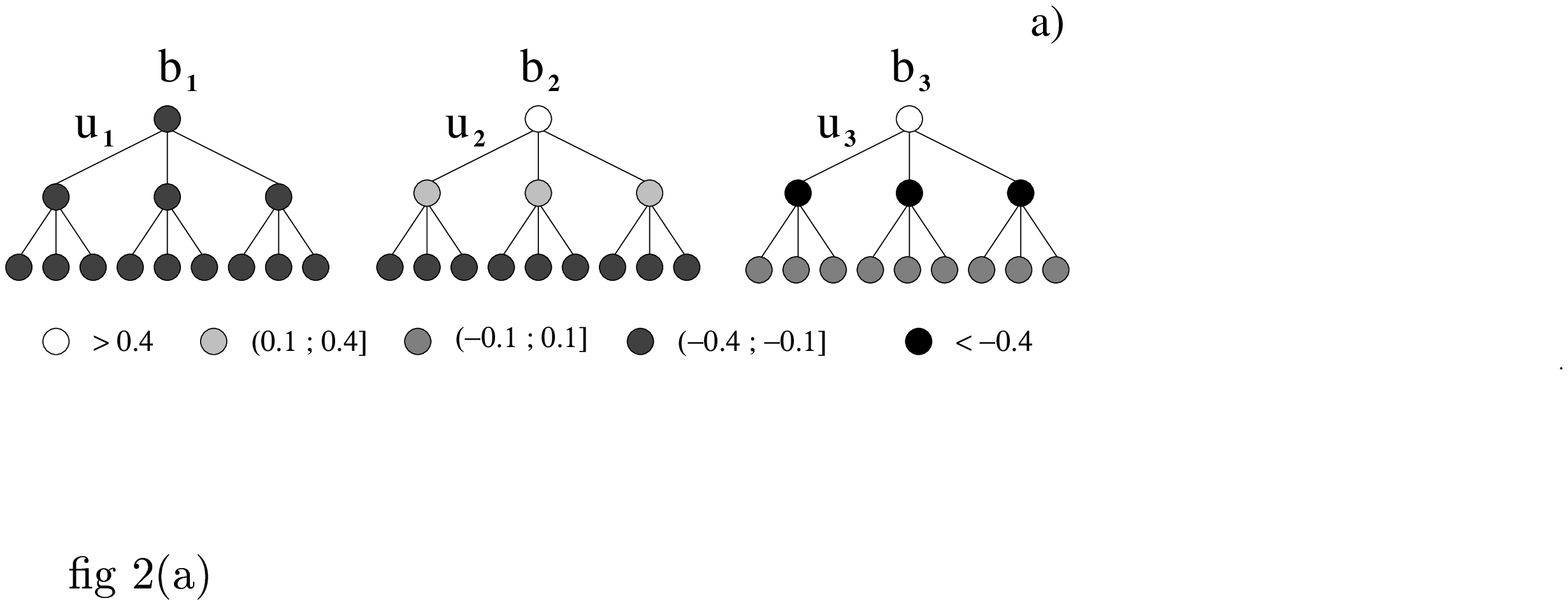, width=0.8\textwidth ,angle=90, clip=}
\epsfig{file=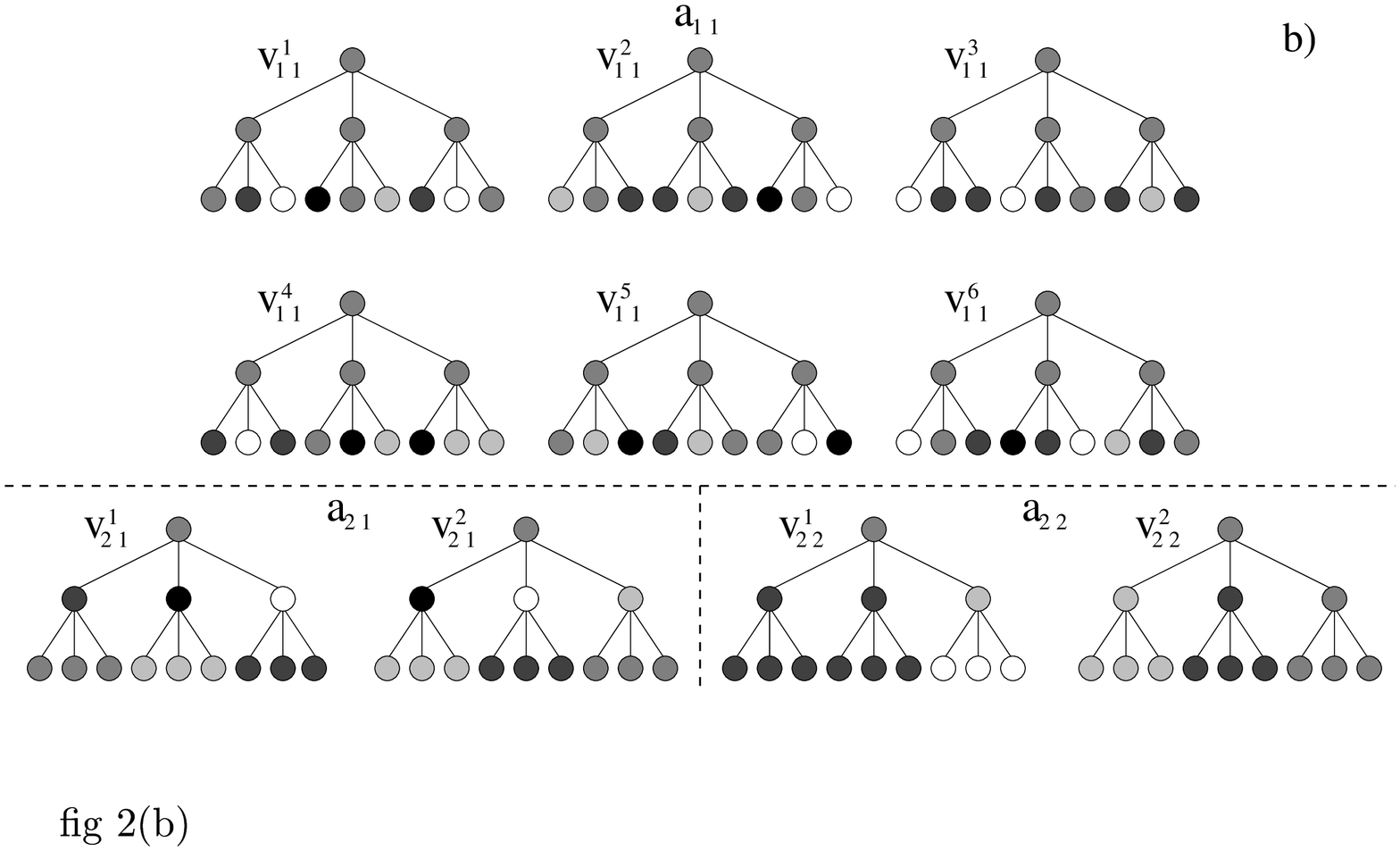, width=0.8\textwidth ,angle=90, clip=}
\caption{ A tree with $R=3$, $L=2$. a) The three non-degenerate, layered eigenvectors
$\{{\bf u}_n\}$ and their corresponding eigenvalues. b) Degenerate eigenvectors
$\{{\bf v}_{m s}^g\}$ and corresponding eigenvalues.}
\end{figure}
\begin{figure}
\epsfig{file=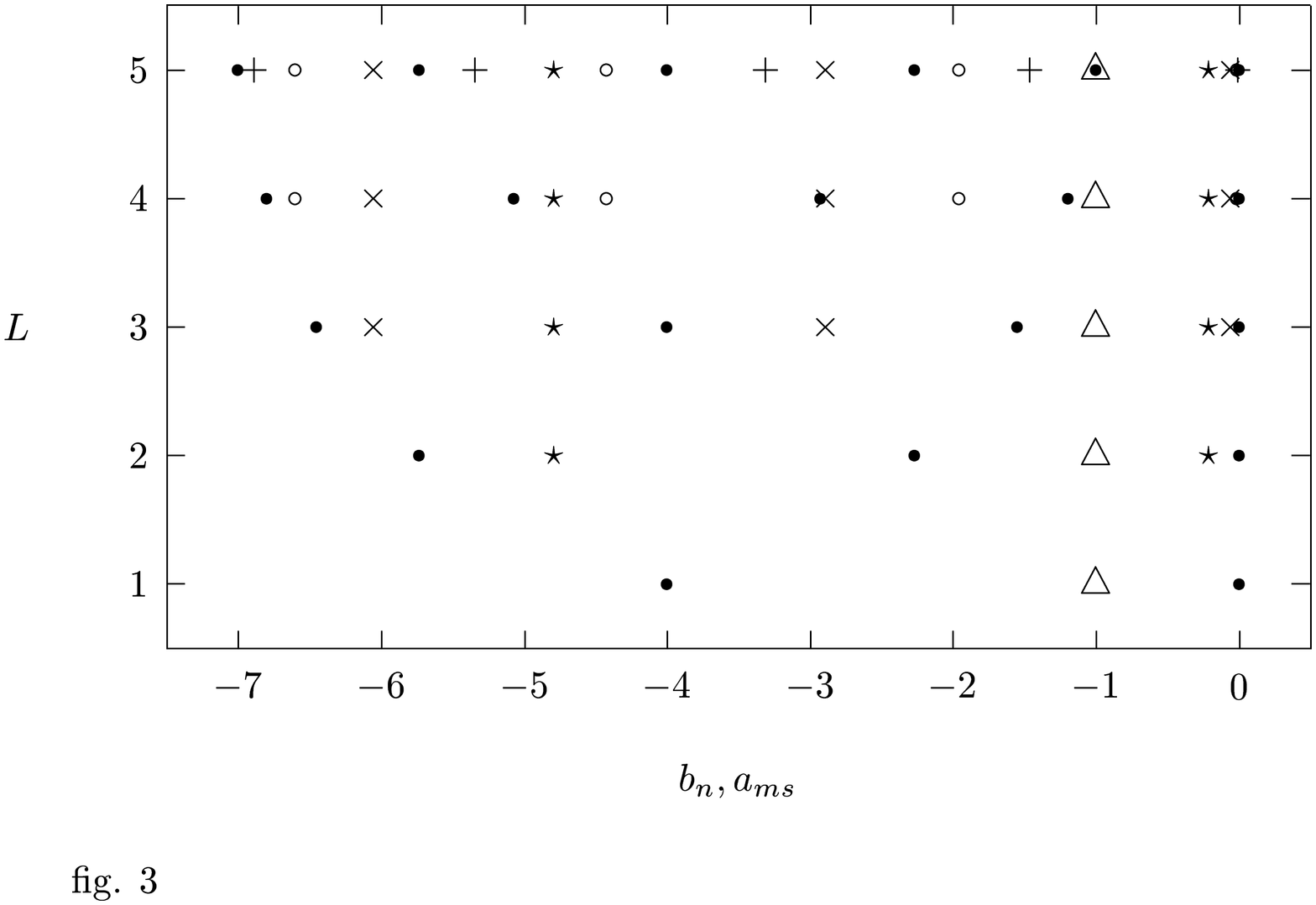, width=1.0\textwidth ,angle=90, clip=}
\caption{Spectrum of eigenvalues at increasing depths $L$, for a tree with
ramification $R=3$. Eigenvalues $\{b_n\}$ are indicated by black dots $(\bullet)$.
Other symbols indicate eigenvalues $\{a_{ms}\}$ as follows:  $a_{1m}$ $(\triangle)$;
$a_{2m}$ $(\star)$; $a_{3m}$ $(\times)$; $a_{4m}$ $(\circ)$; $a_{5m}$ $(+)$.}
\end{figure}
\begin{figure}
\epsfig{file=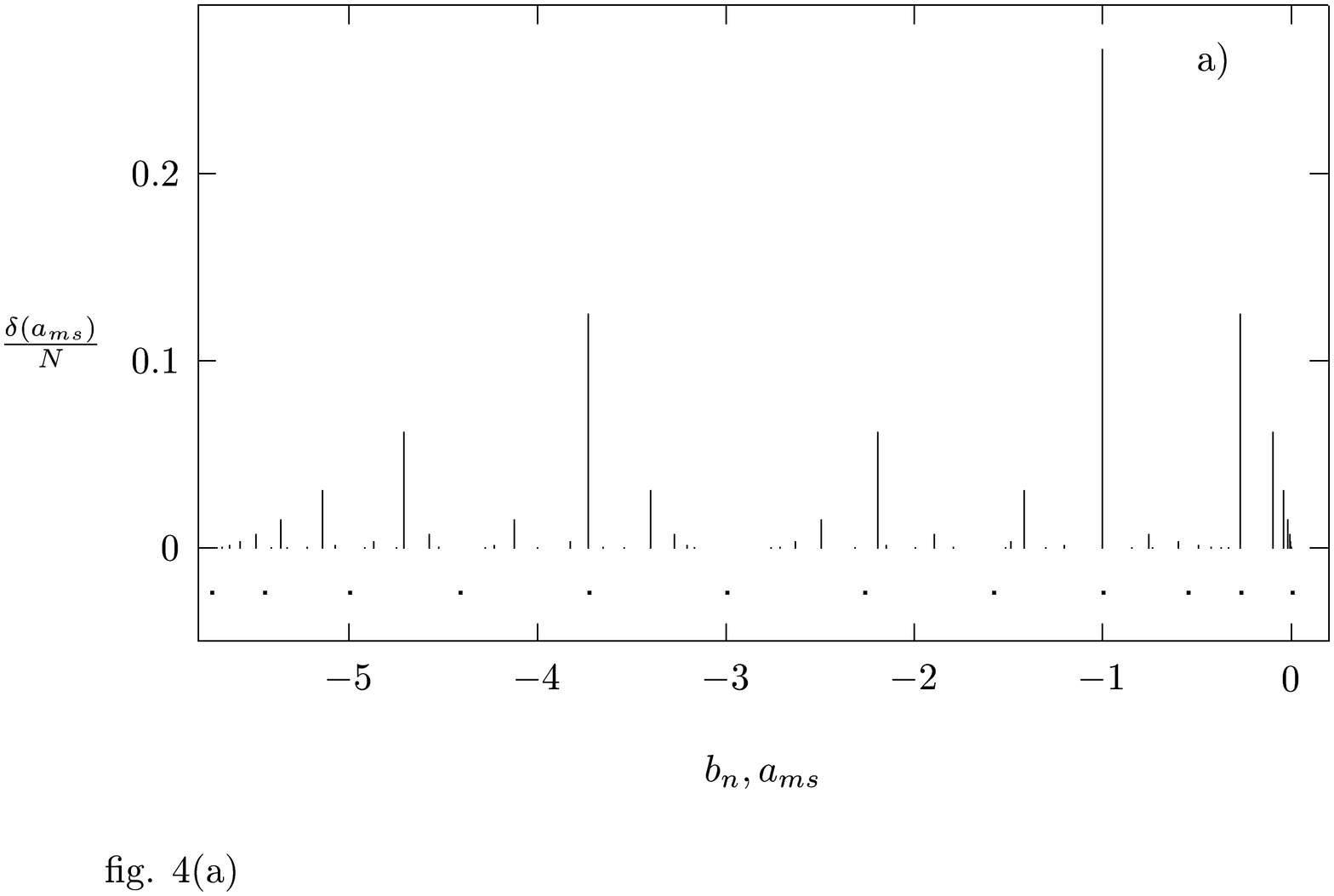, width=0.7\textwidth ,angle=90, clip=}
\epsfig{file=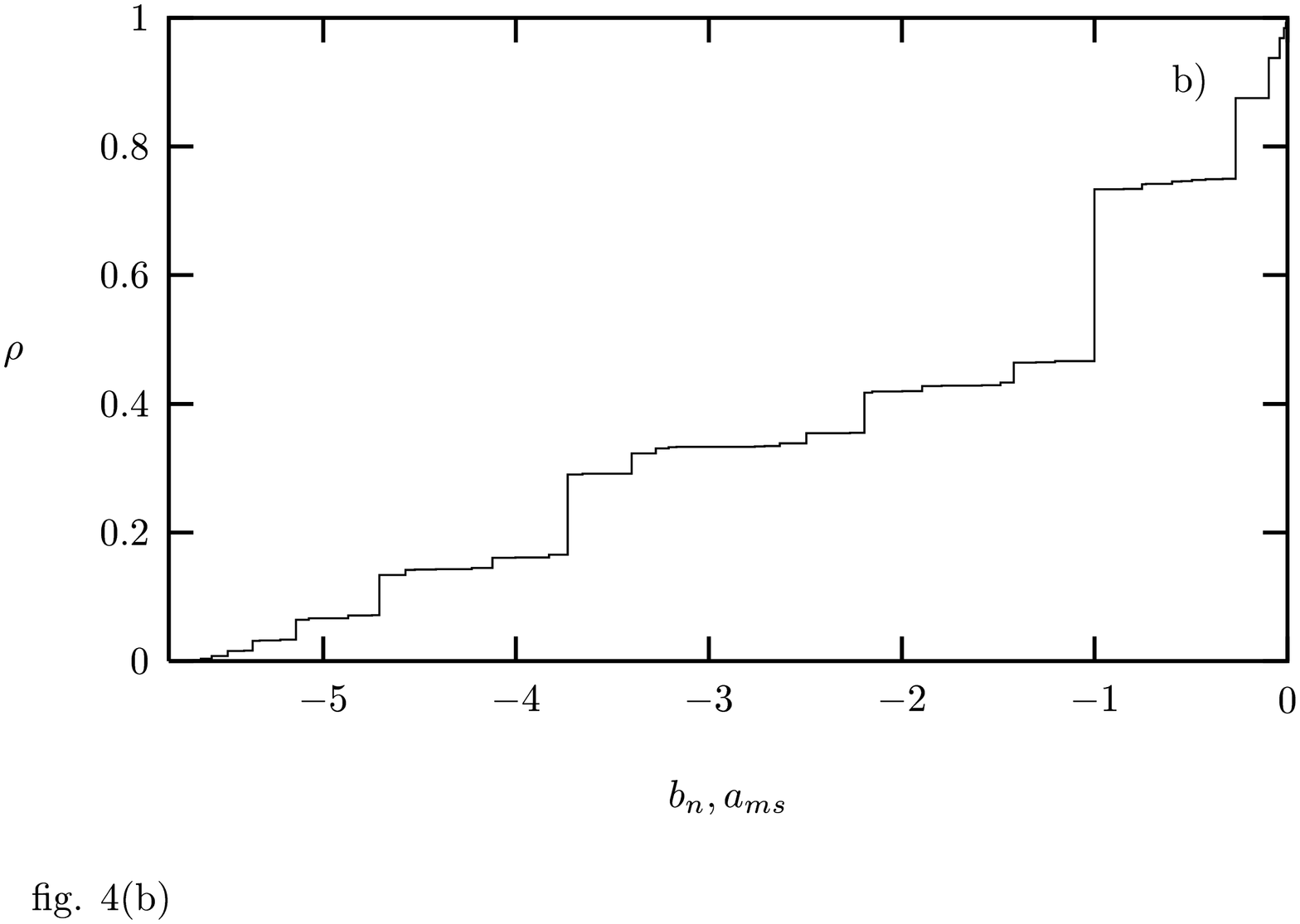, width=0.7\textwidth ,angle=90, clip=}
\caption{a) Full spectrum of eigenvalues of the coupling matrix ${\bf M}$ for a tree
with $R=2$, $L=11$, showing their degeneracy. For clarity, eigenvalues $\{b_n\}$ are
shown with dots just bellow the zero line. The vertical axis shows the degeneracy of
the eigenvalues $\{a_{ms}\}$ divided by $N$, indicated by vertical bars at each
eigenvalue. b) The measure of the set of all eigenvalues of ${\bf M}$.}
\end{figure}
\begin{figure}
\epsfig{file=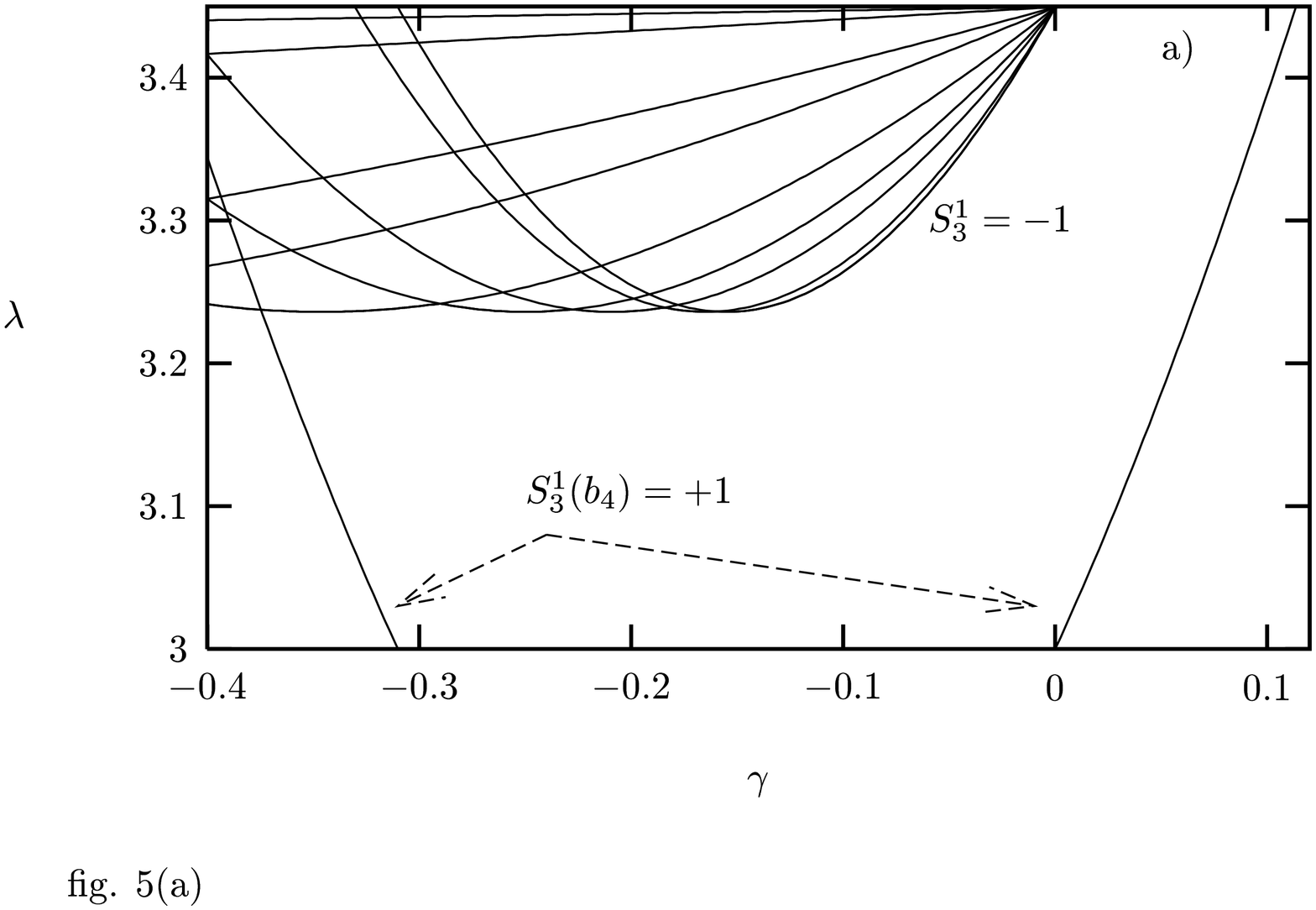, width=0.7\textwidth ,angle=90, clip=}
\epsfig{file=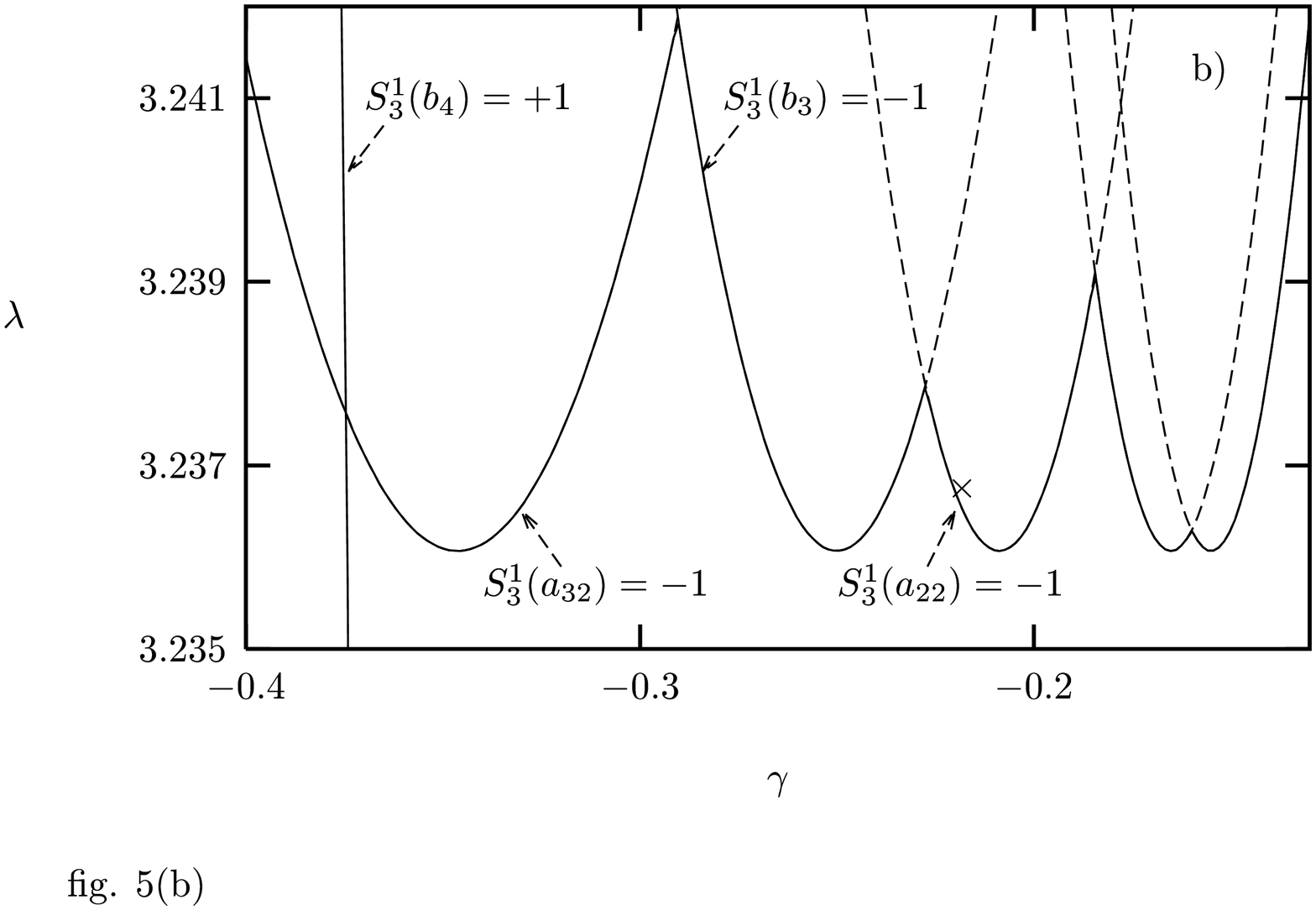, width=0.7\textwidth ,angle=90, clip=}
\caption{a) The boundary curves $S_3^1=\pm1$ given by Eqs.~(\ref{boun1})-(\ref{boun2})
for the period-two, synchronized states of a tree with $R=3$, $L=3$. The curve
$S_3^1(b_4)=+1$ is signaled by arrows. The upper curves correspond to the r.h.s equal
to $-1$ for both types of eigenvalues. The interior region bounded by these curves is
where stable, synchronized, period-two states exist in the plane $(\gamma,\lambda)$.
b) Magnification of the upper curves in a) showing the gaps in the stability boundary
of the period-two, synchronized states. Curves corresponding to several eigenvalues
are indicated by arrows. The cross just beyond the boundary $S_3^1(a_{2\,2})=-1$
indicates the parameters $\gamma$ and $\lambda$ used in Fig.~6.}
\end{figure}
\begin{figure}
\epsfig{file=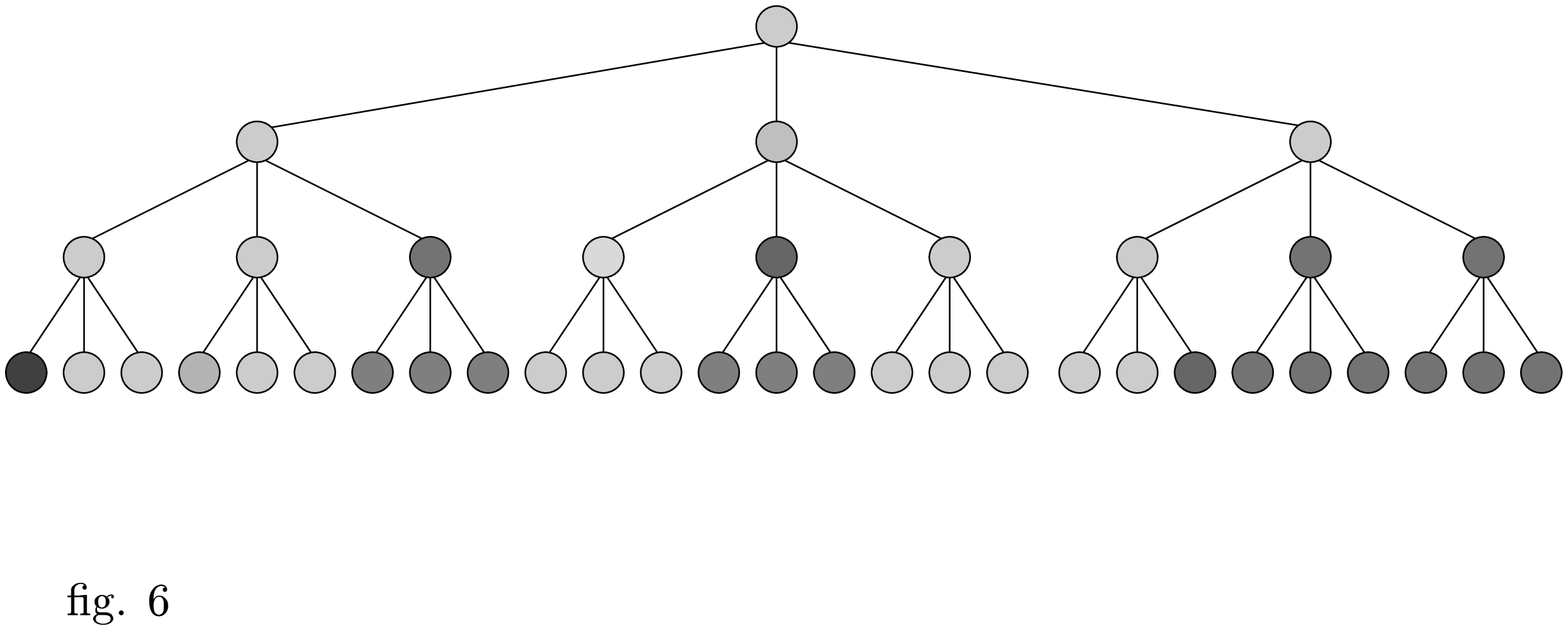, width=1.0\textwidth ,angle=90, clip=}
\caption{Inhomogeneous, period-$4$, state at parameter values $\gamma=-0.22$,
$\lambda=3.2367$ for a tree with $R=3$, $L=3$. This pattern is a linear combination of
the six eigenmodes associated to the eigenvalue $a_{2\,2}$. White corresponds to a
zero value and black to a value equal to one.}
\end{figure}
\end{center}
\end{document}